\documentclass[twocolumn,showpacs,preprintnumbers,amsmath,amssymb,showkeys]{revtex4}

\usepackage{graphicx}
\usepackage{float}
\usepackage{dcolumn}
\usepackage{bm}
\usepackage{color}

\newcommand{\vE}{\bm E}
\newcommand{\vGamma}{\bm \Gamma}
\newcommand{\vtau}{\bm \tau}
\newcommand{\vU}{\bm U}
\newcommand{\vi}{\bm i}
\newcommand{\vn}{\bm n}
\newcommand{\vu}{\bm u}
\newcommand{\vv}{\bm v}
\newcommand{\const}{\operatorname{const}}
\newcommand{\Div}{\operatorname{div}}
\newcommand{\Rcf}{\mathcal{R}}
\newcommand{\zk}{e_k}

\begin{document}

\preprint{APS/123-QED}

\title{Rotating electrohydrodynamic flow in a  suspended liquid film}
\author{E.\;V.~Shiryaeva}
\email{shir@ns.math.rsu.ru}
\affiliation{%
Department of Mathematics, Mechanics and Computer Science, Southern Federal University, 344090, Rostov-on-Don, Russia
}%
\author{V.\;A.~Vladimirov}
 \email{vv500@york.ac.uk}
\affiliation{%
Department of Mathematics, York University, York, YO10 5DD, UK
}%
\author{M.\;Yu.~Zhukov}
\email{zhuk@ns.math.rsu.ru}
\affiliation{%
Department of Mathematics, Mechanics and Computer Science, Southern Federal University, 344090, Rostov-on-Don, Russia
}%

\date{\today}

\begin{abstract}
The mathematical model of a rotating electrohydrodynamic flow in a thin suspended liquid film is proposed and studied.
The motion is driven by the given difference of potentials in one direction and constant external electrical field
$\vE_\text{out}$ in another direction in the plane of a film. To derive the model we employ the spatial averaging over
the normal coordinate to a film that leads to the average Reynolds stress that is proportional to $|\vE_\text{out}|^3$.
This stress generates tangential velocity in the vicinity of the edges of a film that, in turn, causes the rotational
motion of a liquid. The proposed model is aimed to explain the experimental observations of the
\emph{liquid film motor} \cite{Amjadi01,Shirsavar02}.

\end{abstract}

\maketitle

\section{Introduction}

The paper is devoted to the theoretical study of the motions in a thin suspended liquid film. The motions are driven by
the constant external electric field that is applied at the edges of a film. We show that this field produces the
averaged rotating motion of the liquid in the plane of a film. Our aim is to explain the rotating flow observed in an
electrolyze planar water cell placed inside a plane capacitor \cite{Amjadi01, Shirsavar02}. The rotating motion of a
fluid as a whole caused by the action of a \emph{constant} electrical field is so unusual that the authors of
\cite{Amjadi01,Shirsavar02} call this effect a \emph{liquid film motor}, emphasizing that it represents a new type of
engine. They also proposed that it could be explained by the changing of orientation of water molecular dipoles caused
by a strong electric field. Simultaneously they denied the possibility of the generating of such a flow by the edge
effects. In contrast, we show that the jump of an electric field across a water-dielectric interface produces the
tangential velocity of a liquid that can maintain a steady rotating flow in the whole film. In other words, we
demonstrate that one does not need to use a heuristic idea about the switching of the molecular orientations: this
phenomenon can be explained with the use of classical tools only. In our model the rotating motion in a film is
explained by the electro-kinetic effects at its edges. According to our theory the ratio between the spatial scales of
a flow domain plays a crucial role. Only for thin films the classical edge effects can generate the rotation; in
contrast in the flow domains with all spatial scales of the same order this effect will be absent. Naturally, our final
model is two-dimensional (plane); however the tangential velocity at the boundary is actually caused by Reynolds
stresses that appear after the averaging over a film thickness of an original three-dimensional flow. The resulting
tangential velocity has order $O(h^4)$, where $2h$ is a film thickness. An intense electrohydrodynamic (EHD) rotating
flow takes place only in the restricted domain of governing parameters. In particular, such a flow can exist for the
films of moderate thickness (for example, the tangential velocity $\backsim1\,\text{cm}/\text{sec}$ appears for the
following parameters: the strength of the capacitor electric field $\backsim30\,\text{kV}/\text{m}$, the difference of
electrolysis potentials $\backsim20\,\text{V}$, the film thickness $\backsim0.1$--$0.3\,\text{cm}$, and the film
surface size $\backsim1\,\text{cm}$), but it can not exist for very thin films. An important general result of our
paper is the demonstration of the fact that the classical effects (such as the electrokinetic phenomena), that are
small in ordinary conditions, can play the key part in micro-scales. Such revaluation of the classical effects may be
important for the developments of microfluidics and for the creation of microdevices. Here one have take into account
that in this paper we both present systematical theoretical results and show their good agreement with the experiments
of \cite{Amjadi01,Shirsavar02}. The detailed discussion of our results is given in Sect.\,\ref{i:7.0}.

In the mathematical modelling we essentially use the results
\cite{Santiago_001,Santiago_002,Santiago_003,Santiago_004,Santiago_005, Baygentsa_Baldessari}
that contain the analytical and numerical studies of EHD flows with the gradient of conductivity, the method of
depth-average, and an effective asymptotic procedure for the EHD equations of multicomponent mixtures. In the averaged
equations derived in \cite{Santiago_001,Santiago_002,Santiago_003,Santiago_004,Santiago_005} one can see the terms
correspondent to Taylor-Aris dispersion and to Reynolds stresses; however Reynolds stresses are neglected since they
are small for the chosen intervals of parameters and negligible for the studied phenomena. It is also important that in
\cite{Santiago_004} one can  find the comparison between the results for the mathematical models of the different levels of
approximation. The papers, closely related to our studies,
\cite{Basarana,Tilley_Petropoulos_Papageorgioua,Chu_Bazant,Tseluiko,Mortensen} consider EHD flows in thin liquid films
or in liquid layers with interfaces; paper \cite{Vortex_rings} describes the appearance of vortex rings due to
reactions near an electrode; and \cite{Goranovic} presents a rotating EHD flow in a smectic medium. The role of
interface boundary conditions in the electrohydrodynamics (EHD) is well-known from the classical papers
\cite{Melcher_Taylor,Melcher_001, Saville}. The paper
\cite{Zaltzman_Rubinstein} shows that an electrical double layer (EDL)\label{pageEDL1} can allow a slip in
the boundary conditions between a liquid and a solid. The papers
\cite{Bazant_001,Bazant_002,Bazant_003,Ajdari_001,Ajdari_002,Santiago4} are devoted to the influence
of the inhomogeneous electrical charge of microchannel boundaries on EHD flows. Other closely related papers
\cite{Bazant_004,Bazant_005,Bazant_006,Ajdari_003,Ajdari_004} consider various theories of the EDL,
including so-called extremal regimes. The survey of modern EDL-theories can be found in
\cite{Dukhin}.

The interest in various flows of micro- and nano-scales has  increased greatly during the last few years. For example,
the main parts of the surveys \cite{Stone,Squires_Quake} are devoted to EHD processes in microchannels;
\cite{EricksonLi1,ErmakovNano1,ErmakovJacobsonRamsey2000,HuWernerLi,Pennathur_Santiago,Posner} deal with the injection of
a fluid and other flows in microchannels. This interest is strongly stimulated by the creation of the microfabricated
fluid devices for the effective separation or the micromixing of multicomponent mixtures
\cite{KanianskyMasarBodor,BharadwajSantiagoMohammadi,MolhoHerrMosier,JenWuLinWu,JohnsonRossLocascio,OddyMikkelsenSantiago},
the electro-micro-pumps \cite{Micropumps}, \emph{etc}. These new techniques are known as parts of the
\emph{Lab-on-a-Chip technology}.
 The mathematical models in this research area help to understand and to describe micro-processes,
to develop experimental methods, and to construct microchips.

\section{Basic equations}\label{i:2.0}

\begin{figure}[H]
\centering  \includegraphics[scale=0.68]{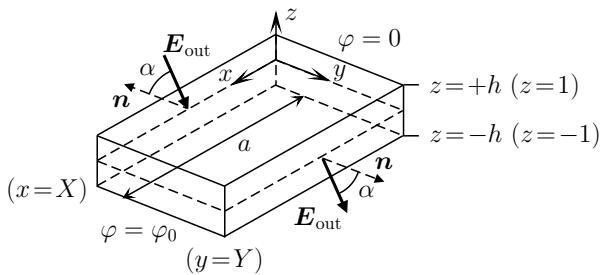}\\
  \caption{A thin film. Dimensionless variables are given in parenthesises.}\label{Ris:SZhV:1}
\end{figure}

The rectangular thin liquid film with the fixed plane free surfaces $z=\pm h$ is considered in Cartesian coordinates
$(x,y,z)$ (Fig.\,\ref{Ris:SZhV:1}). The electrical field can be conveniently split into two parts. The first one is due
to the constant electric potentials $\varphi=0$ and $\varphi=\varphi_0$ on the boundaries $x=0$ and $x=X$, so the
constant difference of potentials is applied in the direction $x$. The second part is the constant external electric
field $\vE_{\text{out}}$ that is prescribed at the boundaries $y=0$ and $y=Y$. The vector $\vE_{\text{out}}$ lies in
the plane $z=\const$, $\alpha$ is the angle between this vector and $y$-axis.
We assume that the electric field is potential, the gravity and the surface tension are absent; the dielectric
permittivity $\varepsilon=\const$ that leads to the absence of the pondermotive force $(1/2)\nabla
\varepsilon (\nabla \varphi)^2=0$. The dimensionless system of governing equations that describes EHD
flows of a multicomponent fluid (for example, water with the ions H$^+$, OH$^-$) is:
\begin{equation}\label{SZhV:eq:1a}
 \delta^2 \frac{d\vu}{dt}
      = -        \delta^2\nabla_0 p+
      \delta^2\nu
        \Delta_0 \vu +      \nu
       \partial_{zz}\vu
      - q\nabla_0 \varphi,
\end{equation}
\begin{equation}\label{SZhV:eq:2a}
 \delta^4 \frac{dw}{dt}
      = -
      \delta^2
      \partial_z  p+  \delta^4 \nu
      \Delta_0 w +   \nu
    \delta^2
      \partial_{zz} w-q\partial_z \varphi,
\end{equation}
\begin{equation}\label{SZhV:eq:3a}
  \Div_0 \vu +  \partial_z w =0,
\end{equation}
\begin{equation}\label{SZhV:eq:4a}
   \varepsilon  \left(
    \delta^2\Delta_0 \varphi +
  \partial_{zz}\varphi\right) =- \delta^2q, \quad   q = \sum_{k}
  \zk c_k,
\end{equation}
\begin{equation}\label{SZhV:eq:5a}
\delta^2\frac{dc_k}{dt}
+ \delta^2 \Div_0 \vi_k+ \partial_z I_k =0,
\end{equation}
\begin{equation}\label{SZhV:eq:6a}
  \vi_k = -D_k(\nabla_0 c_k + \zk \gamma c_k \nabla_0 \varphi),
\end{equation}
\begin{equation*}
 I_k =-D_k(\partial_z c_k + \zk \gamma  c_k \partial_z \varphi),
\end{equation*}
\begin{equation}\label{SZhV:eq:7a}
  \vGamma = (\sigma_{13},\sigma_{23},0)=\nu(\partial_z \vu + \delta^2 \nabla_0 w),
\end{equation}
\begin{equation*}
 \frac{d}{dt}\! =
\partial_t \! + \!
  \vu\! \cdot\! \nabla_0\!  +\!
 w \partial_z,\  \nabla_0\!=(\partial_x,\partial_y),\
 \Delta_0\!=\!\partial_{xx}\!\!+\!\partial_{yy},
\end{equation*}
Here $\vv=(\vu,w)$ and $\vu=(u,v)$ is velocity and its $(x,y)$-projection, $p$ is pressure; $q$ is the density of molar
charge; $\varphi$ is electric potential; $c_k$ is the molar concentration for the $k$-th component of mixture; $\vi_k$
and $I_k$ are the planar and transversal density fluxes for the concentrations; $\nu$ is kinematic viscosity; $D_k$ is
the diffusivity for the components of mixture; $\zk$ are the electric charges of components (in the units of electron's
charge); $\varepsilon$ is the solution permittivity; the parameter $\gamma$ characterises the ratio between the
transports of concentrations by the electrical field and by diffusion; $2\delta$ is a dimensionless film thickness;
$\vGamma$ is the tangential stress vector that is expressed via the components $\sigma_{13}$, $\sigma_{23}$ of a
viscous stress tensor.

On the film boundaries $z=\pm1$ we accept: the no-leak condition for velocity
\begin{equation}\label{SZhV:eq:8a}
  w\bigl|_{z=\pm1}=0,
\end{equation}
the stress-free condition, that with the use of (\ref{SZhV:eq:8a}) is
\begin{equation}\label{SZhV:eq:9a}
  \vGamma\bigl|_{z=\pm1} =\nu(\partial_z \vu + \delta^2 \nabla_0 w)\bigl|_{z=\pm1}=
\nu\partial_z \vu \bigl|_{z=\pm1}=0,
\end{equation}
the no-leak conditions for concentrations
\begin{equation}\label{SZhV:eq:10a}
   I_k\bigl|_{z=\pm1}=0
\end{equation}
and the vanishing of the normal electrical current
\begin{equation}\label{SZhV:eq:11a}
  \partial_z \varphi\bigl|_{z=\pm1} =0.
\end{equation}
We use the governing equations (\ref{SZhV:eq:1a})--(\ref{SZhV:eq:7a}) for the deriving of the averaged model in
Sect.\,\ref{i:3.0} and in Appendix~\ref{app:sec:00}. In our averaging procedure we use only the boundary conditions
(\ref{SZhV:eq:8a})--(\ref{SZhV:eq:11a}). The other boundary conditions (defined for the averaged equations) are given
in Sect.\,\ref{i:3.0}.

For the introducing of dimensionless variables we use the following characteristic values of parameters:
\begin{equation*}
  [x,y] = a,\quad [z] = h, \quad
  [t] = \mathcal{T},  \quad [u,v] = \frac{a}{\mathcal{T}},\quad
    [w] = \frac{h}{\mathcal{T}},\quad
\end{equation*}
\begin{equation*}
  [c_k] = \mathcal{C},\quad
  [E] = \mathcal{E},\quad
  [\varphi]  = \mathcal{E}a,\quad
  [q] =  F\mathcal{C},   \quad \gamma= \frac{F \mathcal{E}a}{RT},\quad
\end{equation*}
\begin{equation}\label{SZhV:eq:12a}
  [p] = F\mathcal{C}\mathcal{E} a\delta^2,\quad
  \mathcal{T}^2=\frac{\rho a}{F\mathcal{C}\mathcal{E}\delta^2},
  \quad \delta^2=\frac{h^2}{a^2}.
\end{equation}
Here $a$ is the characteristic length in the plane of the film; $h$ and $\delta$ are the dimensional and dimensionless
half-thickness of the film; $\rho$ is the density of a liquid; $\mathcal{T}$, $\mathcal{P}$, $\mathcal{C}$ are
characteristic time, pressure, and molar concentration; $F\mathcal{C}$ is characteristic charge density; $F$ is Faraday
constant; $R$ is the universal gas constant; $T$ is the absolute temperature of solution; $a\mathcal{E}$ is the
characteristic difference of electric potentials in the $x$-direction. The dimensional values of kinematic viscosity
$\nu^*$, diffusion coefficients $D_k^*$, and the dielectric permittivity $\varepsilon^*$ are linked to their
dimensionless counterparts as:
\begin{equation}\label{SZhV:eq:13a}
  \nu = \frac{\nu^* \mathcal{T}}{a^2},\quad
  D_k=\frac{D_k^*\mathcal{T}}{a^2},\quad
  \varepsilon= \frac{\varepsilon^* \mathcal{E}}{aF\mathcal{C}}.
\end{equation}
The use of dimensionless parameters (viscosity, diffusivity, \emph{etc.}) instead of conventional scaling numbers
(Reynolds number, Peclet number, \emph{etc.}) is more convenient for our purposes since they allow us to see what
physical effects participate into a certain process. The connections between the introduced parameters and the scaling
numbers are apparent:
\begin{equation*}
  {\rm Re} = \frac{1}{\nu},\quad
  {\rm Pe}_k=\frac{1}{D_k}.
\end{equation*}

\section{The averaging across a film}\label{i:3.0}

The main part of the employed averaging procedure is the same as in
\cite{Santiago_001,Santiago_002,Santiago_003,Santiago_004,Santiago_005}.
The operation of averaging is defined as:
\begin{equation}\label{SZhV:eq:14a}
 \overline{f}  (x,y,t) =
 \frac12 \int\limits_{-1}^1
 f(x,y,z,t)\,dz, \quad \widetilde{f}\equiv f-\overline{f}.
\end{equation}
We decompose the solution of (\ref{SZhV:eq:1a})--(\ref{SZhV:eq:11a}) into the series
\begin{eqnarray}
&\displaystyle\!\!  \{\vu,w,p,q,c_k,\varphi\}=\!\!
    \sum_{m=0}\! \{\vu^m\!,w^m\!,p^m\!,q^m\!,c_k^m\!,\varphi^m\}\delta^{2m}
\!=  \nonumber\\
&\displaystyle = \sum_{m=0}\{\overline{\vu}^m,\overline{w}^m,\overline{p}^m,
  \overline{q}^m,\overline{c}^m_k,\overline{\varphi}^m\}\delta^{2m}+\nonumber\\
   &\displaystyle +\sum_{m=0}\{\widetilde{\vu}^m,\widetilde{w}^m,\widetilde{p}^m,
   \widetilde{q}^m,\widetilde{c}^m_k,\widetilde{\varphi}^m\}\delta^{2m}.
   \label{SZhV:eq:15a}
\end{eqnarray}
The averaging of the governing equations (\ref{SZhV:eq:1a})--(\ref{SZhV:eq:6a}), which takes into account the boundary
conditions (\ref{SZhV:eq:8a})--(\ref{SZhV:eq:11a}) and the decomposition in small parameter $\delta^2$, first yields
$\overline{q}=\overline{q}^0+O(\delta^2)$, $\overline{\varphi}=\overline{\varphi}^0+O(\delta^2)$,
$\overline{c}_k=\overline{c}_k^0+O(\delta^2)$, $\widetilde{\varphi}^0=0$, $\widetilde{c}_k^0=0$, $\widetilde{q}^0=0$
and then leads to the expressions for $\widetilde{\vu}^0$, $\widetilde{w}^0$ è $\widetilde{c}_k^1$. The averaged plane
equations which keep the terms $O(\delta^2)$ are (for the details see Appendix~\ref{app:sec:00})
\begin{eqnarray}
\delta^2\frac{d_0 \overline{\vu}}{dt}   +
\beta\nabla_0(\vU\otimes\vU)
      =  -  \delta^2\nabla_0  \overline{p}+
      \delta^2\nu
        \Delta_0 \overline{\vu}
      - \nu\vU, \label{SZhV:eq:16a}
\end{eqnarray}
\begin{equation}\label{SZhV:eq:17a}
 \Div_0  \overline{\vu}= 0,
\end{equation}
\begin{equation}\label{SZhV:eq:18a}
  \varepsilon
    \Delta_0 \overline{\varphi}  = - \overline{q}, \quad \overline{q} =  \sum_{k}\zk \overline{c}_k,
\end{equation}
\begin{equation}\label{SZhV:eq:19a}
\frac{d_0 \overline{c}_k}{dt}
-\alpha_k\delta^2\Div_0(\vU(\vU \cdot \nabla_0 \overline{c}_k))
+  \Div_0  \overline{\vi}_k = 0,
\end{equation}
\begin{equation*}
\overline{\vi}_k=
-D_k \bigl(\nabla_0 \overline{c}_k + \zk \gamma \overline{c}_k \nabla_0 \overline{\varphi} \bigr),
\end{equation*}
\begin{equation}\label{SZhV:eq:20a}
\frac{d_0}{dt}=\partial_t  +
   \overline{\vu} \cdot \nabla_0, \quad  \nu\vU=\overline{q}\,\nabla_0 \overline{\varphi},
\end{equation}
\begin{equation*}
\beta=\frac{\delta^2}{45},\quad
  \alpha_k=\frac{4}{945D_k},
\end{equation*}
where $(\vU\otimes\vU)$ denotes a tensorial product. We emphasize that after this averaging $\delta$ must be treated as
a regular independent parameter of the problem, together with $\nu$, $\varepsilon$, $D_k$,
\emph{etc.}

For the equations (\ref{SZhV:eq:16a})--(\ref{SZhV:eq:20a}) we prescribe the boundary conditions for the averaged fields
$\overline{\vu}$, $\overline{c_k}$, $\overline{\varphi}$ on the side boundaries $x=0,X$ and $y=0,Y$
(Fig.\,\ref{Ris:SZhV:1}).

The boundaries $y=0$ and $y=Y$ represent the interfaces between two dielectric materials: the liquid with the
dielectric permittivity $\varepsilon$ and the outside medium with the dielectric permittivity
$\varepsilon_{\text{out}}$; these boundaries are insulators (not electrodes), hence we prescribe the continuity of the
normal components for electrical induction \cite{Jackson,Landau}
\begin{equation*}
 \varepsilon (\vn \cdot \nabla_0 \overline{\varphi}) = \varepsilon_{\text{out}}
 (\vn\cdot \vE_{\text{out}}),\quad y=0,\, Y,
\end{equation*}
where $\vn$ is the unit normal vector to the boundary. Since the vector $\vE_{\text{out}}$ lies in the plane $z=\const$
and has the angle $\alpha$ with $y$-axis we have
\begin{equation}\label{SZhV:eq:21a}
\frac{\partial\overline{\varphi}}{\partial \vn} =\pm E_0, \quad y=0,\, Y; \quad E_0=\frac{\varepsilon_{\text{out}}}
{\varepsilon}
|\vE_{\text{out}}|\cos \alpha,
\end{equation}
where the sign `$-$' corresponds to the boundary $y=0$ (Fig.\,\ref{Ris:SZhV:1})
\footnote{~In the case of the insulator-conductor, the interface condition (\ref{SZhV:eq:21a}) correspond
to the surface charge density $\pm\varepsilon_{\text{out}}|\vE_{\text{out}}|\cos\alpha $ on the plates of the capacitor
$y=0,\,Y$. One can also prescribe different boundary condition, for example
$\varphi=\pm\varphi_{\text{out}}\bigl|_{y=0,Y}$.}. The conditions of zero concentration fluxes at $y=0$, $Y$ are
\begin{equation}\label{SZhV:eq:22a}
  \overline{\vi}_k\cdot \vn=0,\quad y=0,Y.
\end{equation}
The fixed difference between the electric potentials at $x=0$, $x=X$ is given as
\begin{equation}\label{SZhV:eq:23a}
 \overline{\varphi}=0,\quad x=0;\quad
 \overline{\varphi}=\varphi_0,\quad x=X.
\end{equation}
For all edge boundaries $x=0$, $X$ and $y=0$, $Y$ we require the no-leak of a liquid
\begin{equation}\label{SZhV:eq:24a}
 \overline{u}\bigl|_{x=0,\,X}=0, \quad
 \overline{v}\bigl|_{y=0,\,Y}=0
\end{equation}
and the conditions
\begin{equation}\label{SZhV:eq:25a}
 \overline{\vu}\cdot \vtau\bigl|_{y=0,\,Y}=-\Rcf
 \nabla_0 \overline{\varphi}\cdot \vtau\bigl|_{y=0,\,Y},
\end{equation}
\begin{equation}\label{SZhV:eq:26a}
 \overline{\vu}\cdot \vtau\bigl|_{x=0,\,X}=-\Rcf
 \nabla_0 \overline{\varphi}\cdot \vtau\bigl|_{x=0,\,X},
\end{equation}
where $\vtau$ is a unit tangent vector to the boundary, $\Rcf$ is the coefficient defined in Sect.\,\ref{i:4.0}. By
virtue of (\ref{SZhV:eq:23a}) the boundary conditions (\ref{SZhV:eq:26a}) for $x=0,X$ take the form of no-slip
condition
\begin{equation}\label{SZhV:eq:27a}
 \overline{v}\bigl|_{x=0,\,X}=0.
\end{equation}
The prescription of the tangential velocity (\ref{SZhV:eq:25a}) at the boundaries $y=0,\,Y$ is justified in
Sect.\,\ref{i:4.0} where we derive it and show that $\Rcf\thicksim E_0^3$. It is derived from the equation
(\ref{SZhV:eq:16a}) that contains the averaged Reynolds stresses
\begin{equation}\label{SZhV:eq:28a}
\beta\nabla_0(\vU\otimes\vU)\equiv\beta (\vU \cdot \nabla_0
 \vU+\vU \Div_0\vU),
\end{equation}
which define $\Rcf$ for certain intervals of parameters $\nu$, $\delta$, $\varepsilon$,
\emph{etc}.

The derivation of (\ref{SZhV:eq:16a})--(\ref{SZhV:eq:20a}) is given in Appendix \ref{app:sec:00}, here we mention only
that the boundary conditions (\ref{SZhV:eq:8a})--(\ref{SZhV:eq:11a}) at $z=\pm1$ play a central part in this
derivation. It is also well-known that if we use only spatial averaging it does not allow us to produce the closed
systems of equations; for its closure one has to employ some additional hypothesis. As such a hypothesis we propose the
condition $\overline{w}^0=0$ that is physically natural and accepted without any mathematical justification. The
equations similar to (\ref{SZhV:eq:16a})--(\ref{SZhV:eq:20a}) have been obtained in
\cite{Santiago_001,Santiago_002,Santiago_003,Santiago_004,Santiago_005} (and other papers cited there) devoted to
the studies of EHD flows with the spatially nonuniform conductivity  in the microchannels with solid boundaries. These
papers also contain the decomposition into the power series and even the term similar to (\ref{SZhV:eq:28a}). However
the key difference with our paper is:  for the physical parameters considered in
\cite{Santiago_001,Santiago_002,Santiago_003,Santiago_004,Santiago_005} this term is small, so it is naturally neglected.

\section{The flows near boundaries}\label{i:4.0}

The problem (\ref{SZhV:eq:16a})--(\ref{SZhV:eq:28a}) can be split into the sequence of two problems: (i) the
calculation of $\Rcf$ in (\ref{SZhV:eq:25a}), and (ii) the finding of the averaged velocity field $\overline{\vu}$ and
the potential $\overline{\varphi}$. In order to evaluate $\Rcf$ we assume that the mixture is  electroneutral
everywhere except the vicinities of the boundaries $y=0,Y$. In these vicinities we build the boundary-layer solution
that leads to a good estimation for $\Rcf$. The detailed studies of the related double layers (the Gouy–Chapman layer,
the Stern layer
\emph{ etc.}) can be found in
\cite{Bazant_002,Bazant_003,Ajdari_001,Ajdari_002,Zaltzman_Rubinstein,Bazant_004,Ajdari_004,Bazant_005,Bazant_006,Dukhin},
where nonlinear and steric effects are taken into account along with the linear electrokinetic effects.\label{pageEDL2} From the
mathematical viewpoint different EDL theories are aimed to formulate and to justify different boundary conditions for
the related boundary layers. The main question is how to choose the mutual positions of a physical boundary and an
interface between the regions with positive and negative charges.

Let us consider the vicinity of the boundary $y=0$ (the case $y=Y$ is similar) and  look for a steady solution of the
problem (\ref{SZhV:eq:16a})--(\ref{SZhV:eq:24a}) neglecting in (\ref{SZhV:eq:19a}) the terms $\alpha_k\delta^2\vU(\vU
\cdot\nabla_0 \overline{c}_k)$
\begin{equation}\label{SZhV:eq:29a}
\overline{\vu}=(\overline{u}(y),0), \quad \overline{c}_k=\overline{c}_k(y), \quad \overline{\varphi}= \Phi(y)+Ex,
\end{equation}
where $E$ is the constant tangential component of the electrical field in the vicinity of $y=0$. The integration of
(\ref{SZhV:eq:19a}) with the boundary conditions (\ref{SZhV:eq:22a}) yields
\begin{equation}\label{SZhV:eq:30a}
 \overline{c}_k(y) = c_{Bk} e^{-\zk \gamma \Phi(y)},\ \
 \overline{q}(y)=\!\sum_k\!  \zk c_{Bk} e^{-\zk \gamma \Phi(y)},
\end{equation}
where $c_{Bk}$ are the constants representing concentrations for the equilibrium Boltzmann distributions. We restrict
ourselves with the case when the mixture is electroneutral, only two kinds of ions are present  (for example
$\mathrm{H}^+$ and $\mathrm{OH}^-$ for water), and the equilibrium Boltzmann distribution is valid:
\begin{equation}\label{SZhV:eq:31a}
  c_{B1}= c_{B2} \stackrel{\text{def}}{=} c_{B},\quad
  z_1=1,\quad z_2=-1.
\end{equation}
The Poisson-Boltzmann equation (\ref{SZhV:eq:18a}) takes the form
\begin{equation}\label{SZhV:eq:32a} 
  \lambda^2\partial_{yy}\theta=\sinh\theta, \quad \theta(y)=\gamma\Phi(y).
\end{equation}
where $\lambda_D$ (or $\lambda$)  are (or relative) Debye's length:
\begin{equation}\label{SZhV:eq:33a} 
  \lambda^2=\frac{\varepsilon}{2\gamma c_B}=\frac{\lambda_D^2}{a^2}\ll 1, \quad
  \lambda_D^2=\frac{\varepsilon^* RT }{2c_B^*F^2 }.
\end{equation}
In the vicinity of $y=0$ the boundary layer variable is introduced as $y=\lambda \eta$  (similarly, at $y=Y$ the change
of variable is $y=Y+\lambda \eta$). In the more precise terms the considered
boundary-layer solution represents so called `penetrating boundary layer' \cite{Ilin}. In this case the original
equations and the boundary-layer equations coincide, and $\lambda\ll 1$ is not required for the obtaining of the
solution. Instead of the looking for the boundary-layer solution decaying at infinity one can use the symmetry with
respect to the middle of the domain ($y=Y/2$); the result will remain the same. Nevertheless our further consideration
follows the path that is more transparent from the physical viewpoint. The equation (\ref{SZhV:eq:32a}) takes form
\begin{equation}\label{SZhV:eq:34a} 
  \partial_{\eta\eta}\theta(\eta)=\sinh\theta(\eta).
\end{equation}
Its integration with the boundary condition (\ref{SZhV:eq:21a}) for $\Phi(0)$ yields
\begin{equation}\label{SZhV:eq:35a} 
\theta(0)=-\theta_0, \quad \gamma\Phi(0)=-\theta_0
\end{equation}
where
\begin{equation*}
  \theta_0=\ln\left(1+\mathbb{E}^2+\mathbb{E}\sqrt{2+\mathbb{E}^2}\right),\quad
  \mathbb{E}^2=\frac{\gamma \varepsilon}{4c_B} E_0^2,
\end{equation*}
The expression for $\Phi(Y)$ is similar to (\ref{SZhV:eq:35a})
\begin{equation*}
\theta(0)=\theta_0,  \quad \gamma\Phi(Y)=\theta_0,
\end{equation*}
The boundaries $y=0$ and $y=Y$ represent the different plates of the capacitor, therefore the opposite signs of the
potential $\Phi$ are apparent. The calculation of $\beta\nabla_0(\vU\otimes\vU)$ with the use of (\ref{SZhV:eq:28a}),
(\ref{SZhV:eq:20a}), (\ref{SZhV:eq:29a}) yields
\begin{equation}\label{SZhV:eq:36a}
\beta\nabla_0(\vU\otimes\vU) =
 \frac{\beta}{\nu^2} \left(
E \partial_y \left(
 \overline{q}^2  \partial_y \Phi\right), \partial_y
  \left( \overline{q}\partial_y \Phi\right)^2
 \right).
\end{equation}
where the righthand side (written in components) allows us to integrate the equation (\ref{SZhV:eq:16a}) with the
additional condition $u(\infty)=0$. This condition means that the flow arising near the boundary must decay at large
distances, \emph{i.e.} that the distributions of the horizontal component $u$ and potential $\Phi$ are of a
boundary-layer type (for more details see Appendix~\ref{app:sec:01})
\begin{equation}\label{SZhV:eq:37a}
  \frac{\beta E \varepsilon^2}{\nu^2 \lambda^4 \gamma^3}
\int\limits_\infty^\eta
 (\partial_{\eta \eta}  \theta)^2 \partial_\eta \theta\,d\eta
=       \delta^2\nu
         u(\eta)
      + \frac{\varepsilon E \theta(\eta)}{\gamma}.
\end{equation}
Recall once again that the solutions for $u$ and  $\Phi$ represent a `penetrating boundary layer', so one can obtain the exact
solution with the use of symmetry by taking $u=0$ at the middle of the domain ($y=Y/2$). For the obtaining of the
boundary condition (\ref{SZhV:eq:25a}) and defining $\Rcf$ we evaluate the integral at $\eta=0$:
\begin{equation*}
\int\limits_\infty^0
 (\partial_{\eta \eta}  \theta)^2 \partial_\eta \theta\,d\eta
=\int\limits_0^{\theta(0)}
 \sinh^2 \theta\,d\theta=\frac12(\sinh \theta_0\cosh \theta_0-\theta_0).
\end{equation*}
Taking into account that $E=-(\vtau\cdot \nabla_0 \varphi)_{y=0}$ and comparing (\ref{SZhV:eq:37a}) with
(\ref{SZhV:eq:25a}) we obtain (the case $y=Y$ is similar)
\begin{equation}\label{SZhV:eq:38a} 
\Rcf_3 =\pm   \frac{2c_B^2\beta }{\delta^2\nu^3\gamma} (\sinh \theta_0\cosh \theta_0-\theta_0), \quad
 \Rcf_1 = \pm
 \frac{\varepsilon}{\delta^2\nu\,\gamma}
   \theta_0,
\end{equation}
\begin{equation*}
\Rcf=\Rcf_3+\Rcf_1,
\end{equation*}
where different signs correspond to $y=0$ and $y=Y$. With a sufficient precision the value of $\Rcf_3$ at
$\mathbb{E}\leqslant 1.5$ can be replaced with
\begin{equation}\label{SZhV:eq:39a} 
 \Rcf_3 \approx \pm  \frac{8c_B^2\beta } {3\delta^2\nu^3\gamma}
  \sqrt{2}\,\mathbb{E}^3,\quad  \mathbb{E}\leqslant 1.5.
\end{equation}
In order to avoid misunderstanding we should mention that the calculated value of $\Rcf$ (\ref{SZhV:eq:25a}) represents
only a rough estimation; to obtain it we accept that the equation (\ref{SZhV:eq:19a}) is steady and neglect the
Taylor-Aris dispersion. Moreover, in (\ref{SZhV:eq:29a}) we assume that $E=\const$ on the boundaries $y=0,Y$ that is
not true, later on we consider $E=E(x)$ (see\,(\ref{SZhV:eq:45a})).
 \label{page11} In spite of these simplifying assumptions, the
results of this section show that Reynolds stresses $\beta\nabla_0(\vU\otimes\vU)$ for certain parameters can crucially
participate to the generation of the tangential velocity (of order $O(E^3_0)$) at the side boundary of a film.

\section{The flow in a thin film}\label{i:5.0}

In order to describe the flow in a thin film we use the simplified version of the equations
(\ref{SZhV:eq:16a})--(\ref{SZhV:eq:20a}), where we accept that the mixture is electroneutral ($\overline{q}=0$)
everywhere but the vicinities of the boundaries. It allows us to eliminate from the equations all terms proportional to
$\vU$, taking them into account only in the boundary conditions (see~Sect.\,\ref{i:4.0}). The problem describing the
averaged velocity $\overline{\vu}=(\overline{u},\overline{v})$ and the averaged potential $\overline{\varphi}$ is
\begin{equation}\label{SZhV:eq:40a} 
\partial_t \overline{\vu}  +\overline{\vu} \cdot \nabla_0 \overline{\vu}
= - \nabla_0 \overline{p}+ \nu  \Delta_0 \overline{\vu},\quad
  \Div_0 \overline{\vu}= 0.
\end{equation}
\begin{equation}\label{SZhV:eq:41a}
  \Delta_0 \overline{\varphi} =0,
\end{equation}
where (\ref{SZhV:eq:41a}) corresponds to the continuity equation for an electric current in the case of constant
conductivity and equal diffusion coefficients $D_k$ (see Appendix~\ref{app:sec:02});  (\ref{SZhV:eq:41a}) is not the
Poisson-Boltzmann equation (\ref{SZhV:eq:18a}) that express the potential via the charge that was used in
Sect.\,\ref{i:4.0}, the formal coincidence of these equations should not cause misunderstanding. We solve
(\ref{SZhV:eq:40a}), (\ref{SZhV:eq:41a}) in the rectangular domain $D=\{ 0\leqslant x\leqslant X,\ 0\leqslant
y\leqslant Y\}$ with the boundary conditions (\ref{SZhV:eq:21a}), (\ref{SZhV:eq:23a}), (\ref{SZhV:eq:24a}),
(\ref{SZhV:eq:26a}), (\ref{SZhV:eq:27a})
\begin{equation}\label{SZhV:eq:42a}
 \overline{\vu}\bigl|_{x=0,\, X} =0,\quad
 \overline{v}\bigl|_{y=0,\, Y} =0
\end{equation}
\begin{equation}\label{SZhV:eq:43a}
 \overline{u}\bigl|_{y=0,\,Y} = -\Rcf\overline{\varphi}_x\bigl|_{y=0,\,Y},\quad
 \Rcf = \Rcf_1(E_0) + \Rcf_3(E_0),
\end{equation}
\begin{equation}\label{SZhV:eq:44a}
 \overline{\varphi}\bigl|_{x=0} =0,\ \
\overline{\varphi}\bigl|_{x=X} =\varphi_0,\ \
 \overline{\varphi}_y\bigl|_{y=0,\,Y} =E_0.
\end{equation}
Recall that the expression for $\Rcf$ is given by (\ref{SZhV:eq:38a}), (\ref{SZhV:eq:39a}) and the value of $\Rcf$
essentially depends on $E_0$.

\begin{figure}[H]
\centering  \includegraphics[scale=0.65]{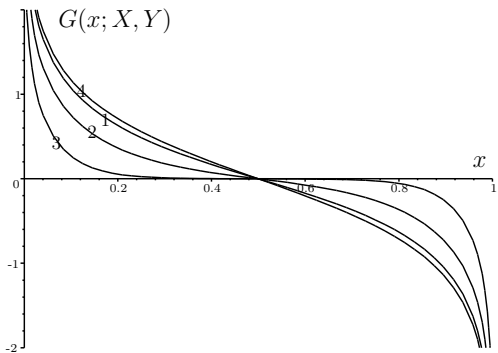}\\
  \caption{The function $G(x;X,Y)$ for $X=1$ and the different values of $Y$:
   (1) $Y=1$; (2) $Y=0.5$; (3) $Y=0.2$; and (4) $Y=2$}\label{Ris:SZhV:2}
\end{figure}

The problem (\ref{SZhV:eq:41a}), (\ref{SZhV:eq:44a}) has an analytic solution that can be presented as Fourier's
series. For the further use we give only the following formula (where the sign `$+$' corresponds to $y=0$)
\begin{equation}\label{SZhV:eq:45a}
  \overline{\varphi}_x\bigl|_{y=0,\,Y}  =\frac{\varphi_0}{X}\pm
  E_0G(x;X,Y),
\end{equation}
\begin{equation*}
G(x;X,Y)=\frac{4}{\pi} \sum_{k=0}^\infty \frac{\tanh \frac{(2k+1)\pi Y}{2X}}{(2k+1)} \cos(2k+1)\frac{\pi x}{X}.
\end{equation*}
The computed graphs of $G(x;X,Y)$ for different values of $X$, $Y$ are shown in Fig.\,\ref{Ris:SZhV:2}.
\noindent
It is apparent that for the fixed $X$, $Y$ the sign of $\overline{\varphi}_x\bigl|_{y=0,\,Y}$ (and hence the tangential
velocity $\overline{u}\bigl|_{y=0,\,Y}$ given by (\ref{SZhV:eq:43a})) depends on the relation between the parameters
$\varphi_0$, $E_0$. For example, for $X=Y=1$ and $\varphi_0/X=E_0$ the velocity $u<0$ on the part of the boundary
$\{0<x\lesssim 0.1,y=0\}$, and $u>0$ on the rest of it $\{0.1 \lesssim x<1,y=0\}$.

\section{The Numerical Results}\label{i:6.0}

We solve the Navier-Stokes equations (\ref{SZhV:eq:40a}) with the prescribed tangential velocity and the no-leak
condition at $y=0,\,Y$ and the no-slip condition at the rest of the boundary (\ref{SZhV:eq:42a}), (\ref{SZhV:eq:43a})
by the employment of the standard projection algorithm \cite{Chorin,Rannacher} and the finite element method. The
numerical setting is based on the package FreeFem++ \cite{RukoFFp} with the use of adaptive grids. The formula
(\ref{SZhV:eq:45a}) for $\overline{\varphi}_x\bigl|_{y=0,\,Y}$  is not efficient due to its singularities at $x=0,\,X$;
therefore taking into account the singularities of derivatives near the vertices we also find $\overline{\varphi}$
(\ref{SZhV:eq:41a}), (\ref{SZhV:eq:44a}) numerically.

\begin{figure}[H] 
\centering  \includegraphics[scale=0.85]{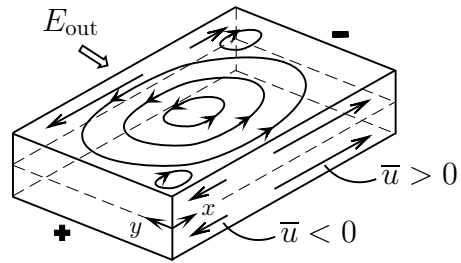}\\
  \caption{The sketch of a rotating flow in the film}\label{Ris:SZhV:3}
\end{figure}

The formulated problem is rather simple, however the qualitative properties of its solution strongly depend on the
relation between the parameters $\varphi_0$, $E_0$, $X$, $Y$. As we have already mentioned  the direction of the
tangential velocity on the boundaries $y=0,\,Y$ is defined by (\ref{SZhV:eq:45a}) (Fig.\,\ref{Ris:SZhV:2}): the
velocity is positive on one part of the boundary and negative on its remaining  part  (Fig.\,\ref{Ris:SZhV:3}) in such
a way that the particular velocity distribution depends mainly on the ratio $\varphi_0/E_0$. It is apparent that this
tangential velocity causes the rotational motion of a large scale. Additional smaller vortices can appear in the
regions adjacent to the parts of the boundary, where the tangential velocity has the opposite sign
(Fig.\,\ref{Ris:SZhV:3}).
It is instructive to express dimensionless parameters in terms of dimensional ones with the use of (\ref{SZhV:eq:12a}),
(\ref{SZhV:eq:13a}), (\ref{SZhV:eq:38a}), (\ref{SZhV:eq:39a})
\begin{equation}\label{SZhV:eq:46a}
  \frac{a}{\mathcal{T}}\Rcf_1 = -
 \frac{\varepsilon^*\mathcal{E}^2}{\rho\nu^*}
\lambda_D      E_0,
\end{equation}
\[
   \frac{a}{\mathcal{T}}\Rcf_3\approx   \frac{F c_B^*\varepsilon^*\mathcal{E}^3 } {135\rho^2{\nu^*}^3}
  \left(\frac{2 \varepsilon^* \mathcal{E}^2 }{ RT c_B^*} \right)^{1/2}\!\!\!\!
  E_0^3
  h^4 ,
\]
\[
\lambda_D=\left(\frac{\varepsilon^* RT }{2c_B^*F^2 } \right)^{1/2}\!\!\!\!,\quad
    \left(\frac{\varepsilon^*  \mathcal{E}^2  }{4 RT c_B^*} \right)^{1/2}\!\!\!\!
     E_0\leqslant 1.5.
\]
\[
 \mathcal{T} = \frac{1}{\delta}\sqrt{\frac{\rho a}{F\mathcal{C}\mathcal{E}}},\ \
  \nu = \frac{\nu^* \mathcal{T}}{a^2},\ \
  E_0=\frac{\varepsilon_\text{out}^*}{\varepsilon^*}
  \frac{E_{\text{out}}^*}{\mathcal{E}},
  \ \  \varphi_0=\frac{\varphi_0^*}{a\mathcal{E}}.
\]
We perform our computations for the experimental values of \emph{parameters for a liquid film motor} taken from
\cite{Amjadi01,Shirsavar02}; all used values are listed in Tables~\ref{Tabl:SZhV:1}, \ref{Tabl:SZhV:2},
 \ref{Tabl:SZhV:3}. It is apparent that the velocity $(a/\mathcal{T})\Rcf_1$ (that is similar to the classic
 electroosmosis)
is significantly less than the tangential velocity on the boundary $( a/\mathcal{T})\Rcf_3$ that appears due to the
averaging over the film thickness. Therefore in the computations we have not taken $\Rcf_1$ into account.
\begin{table}[H]
\centering
\caption{Dimensional parameters}\label{Tabl:SZhV:1}
\noindent \begin{tabular}{|c|l|l|}
\hline
 \multicolumn{1}{|c|}{Symbol} &
 \multicolumn{1}{c|}{Description} &
 \multicolumn{1}{c|}{Value} \\
\hline
 $\varphi_0^*$ &  difference of potentials&  20\,\textrm{V} \\
 $a$ &length  & $10^{-2}\,\textrm{m}$ \\
 $E_{\text{out}}^*$ & electric intensity & $30000\,\textrm{V}/\,\textrm{m}$  \\
 $\nu^*$ & kinematic viscosity & $10^{-6}\,\textrm{m}^2/\textrm{s}$ \\
$\varepsilon^*_0$ & absolute permittivity  & $8.85\cdot 10^{-12}\,\textrm{C}/(\textrm{V}\cdot\textrm{m})$ \\
$\varepsilon^*$ & water permittivity  & $78.3\,\varepsilon^*_0$ \\
$\varepsilon_{\text{out}}^*$ & air permittivity   & $1.0\,\varepsilon^*_0$  \\
 $\rho$ & water density  &  $10^3\,\textrm{kg}/\textrm{m}^3$ \\
 $\mathcal{C}=c_B^*$ & ion concentration & $10^{-4}\,\textrm{mol}/\textrm{m}^3$ \\
 $F$ & Faraday constant & $9.65\cdot 10^{4}\,\textrm{C}/\textrm{mol}$\\
 $R$ & universal gas constant & $8.3\,\textrm{J}/(\textrm{mol}\cdot\textrm{K})$ \\
 $T$ & absolute temperature & $293\,\textrm{K}$ \\
\hline
\end{tabular}
\end{table}

\begin{table}[H]
\centering
\caption{Characteristic scales}\label{Tabl:SZhV:2}
\noindent \begin{tabular}{|c|l|l|}
\hline
 \multicolumn{1}{|c|}{Symbol} &
 \multicolumn{1}{c|}{Description} &
 \multicolumn{1}{c|}{Value} \\
\hline
 $\mathcal{E}$ & electric strengths scale & $2000\,\textrm{V}/\textrm{m}$ \\
 $\mathcal{T}$ & time scale & $7.8\cdot 10^{-2}\,\textrm{s}$ \\
 $a/\mathcal{T}$ & velocity scale & $0.128\,\textrm{m}/\textrm{s}$ \\
 $\Rcf_3( a/\mathcal{T})$ & tangent velocity scale & $3\cdot 10^{-2}\,\textrm{m}/\textrm{s}$ \\
  $\Rcf_1( a/\mathcal{T})$ & tangent velocity scale &  $0.5\cdot 10^{-6}\,\textrm{m}/\textrm{s}$ \\
  $\lambda_D$ & Debye's length & $0.95\cdot 10^{-6}\,\textrm{m}$\\
  $h=\delta a$ & halfheight & $0.29\cdot 10^{-2}\,\textrm{m}$\\
  \hline
\end{tabular}
\end{table}

\begin{table}[H]
\centering
\caption{Dimensionless parameters}\label{Tabl:SZhV:3}
\noindent \begin{tabular}{|c|c|c|c|c|c|c|c|c|}
\hline
 Fig. & $E_0$ & $\varphi_0$ & $\delta$ & $\nu$ &  $\Rcf_3$ &  $\Rcf_3/E_0^3$ &  $X$ &  $Y$ \\
  \hline
  \ref{Ris:SZhV:Phi_norma}--\ref{Ris:SZhV:uv}, \ref{Ris:SZhV:psi_round}
  & $0.19$ & $-1.0     $  & $0.29$ & $7.8\cdot 10^{-4}$ &  $0.235$ &  $33.42$ &  $1.0$ &  $1.0$ \\
   \hline
 \ref{Ris:SZhV:psi21} & $0.19$ & $-1.0     $  & $0.29$ & $7.8\cdot 10^{-4}$ &  $0.235$ &  $33.42$ &  $1.0$ &  $0.5$ \\
   \hline
\ref{Ris:SZhV:9} & $0.19$ & $-0.1     $  & $0.29$ & $7.8\cdot 10^{-4}$ &  $0.235$ &  $33.42$ &  $1.0$ &  $1.0$ \\
   \hline
\end{tabular}
\end{table}
One can see that $\delta^2\approx 0.09$; it gives us a sufficient ground to treat $\delta^2$ as a small parameter and
to use (\ref{SZhV:eq:16a})--(\ref{SZhV:eq:20a}).

\begin{figure}[H]
\centering  \includegraphics[scale=0.5]{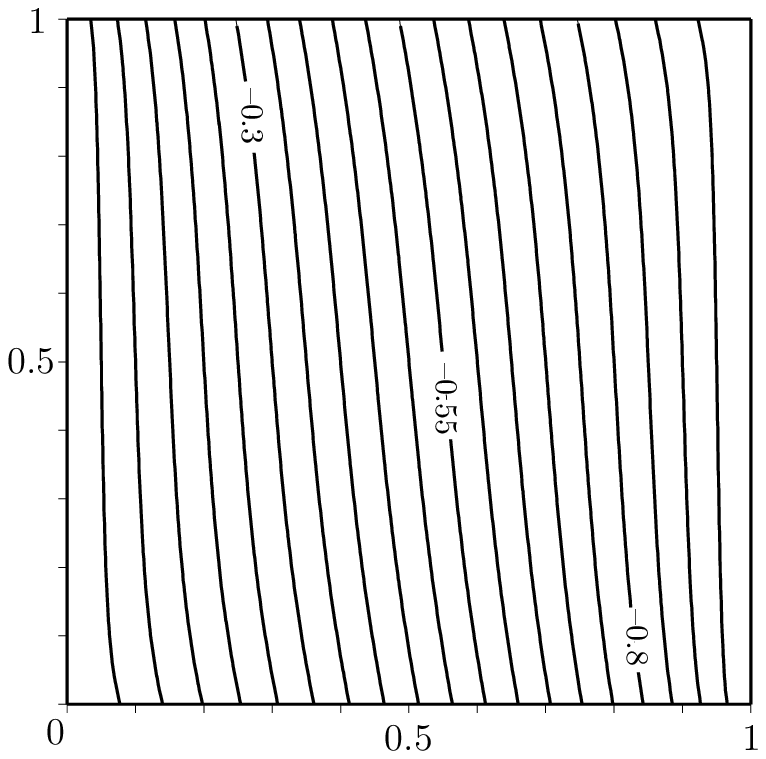}\quad
\centering  \includegraphics[scale=0.65]{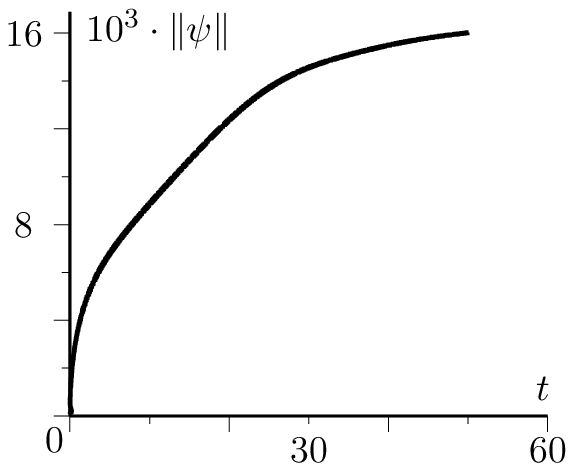}\\
  \caption{The isolines of the potential $\overline{\varphi}(x,y)$ (left) and $\|\overline{\psi}(\cdot,t)\|$}\label{Ris:SZhV:Phi_norma}
\end{figure}

\begin{figure}[H]
\centering
\includegraphics[scale=0.5]{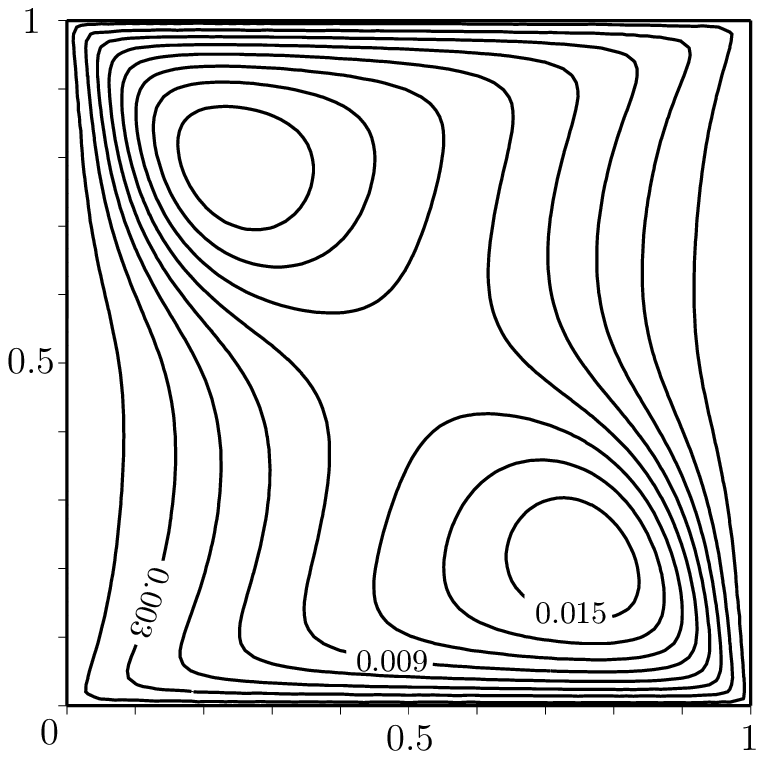}\quad
\includegraphics[scale=0.5]{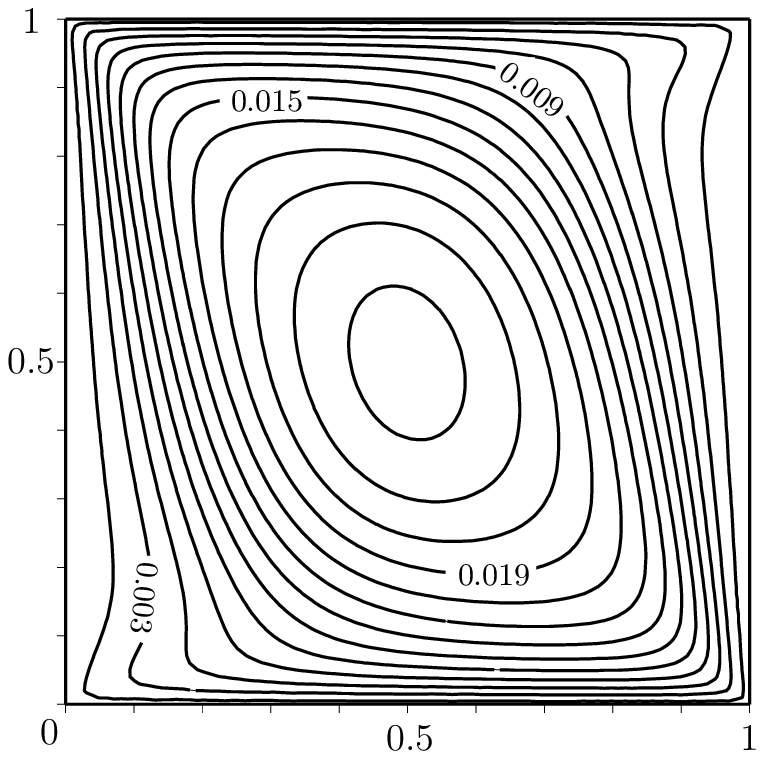}\\
  \caption{The streamlines of $\overline{\psi}(x,y,t)$ for $t=10$ ($\thickapprox 0.78~s$)
  and $t=30$ ($\thickapprox 2.34~s$)}\label{Ris:SZhV:psi}
\end{figure}

\begin{figure}[H]
\centering
\includegraphics[scale=0.50]{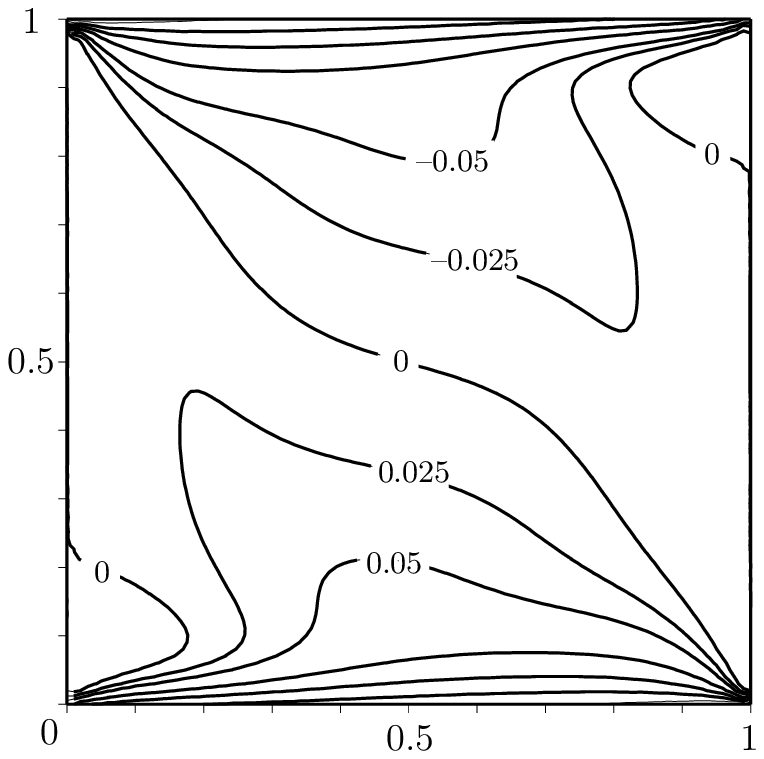}\quad
\includegraphics[scale=0.50]{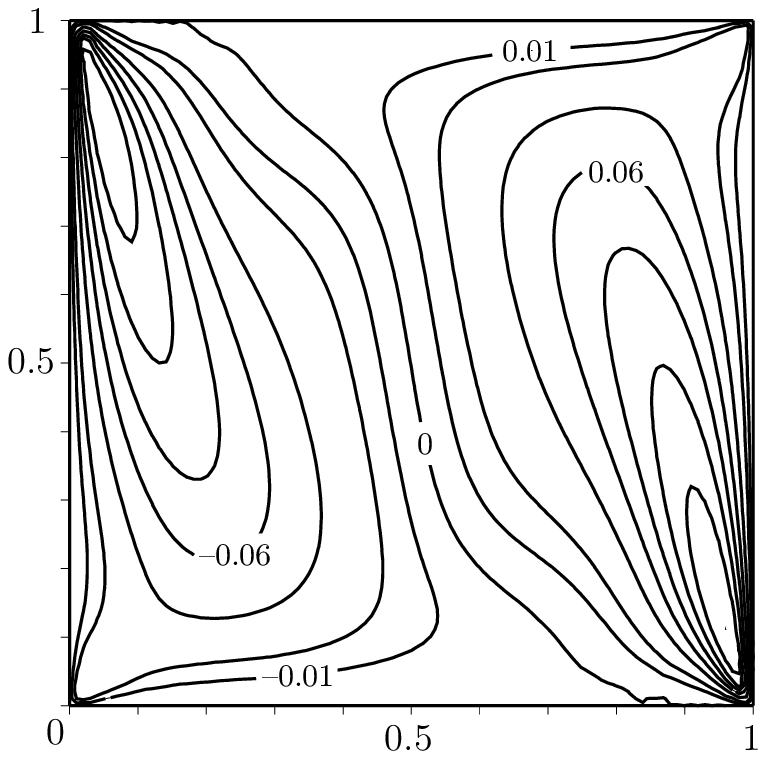}\\
  \caption{The isolines of $\overline{u}(x,y,t)$ (left) and $\overline{v}(x,y,t)$ at $t=30$}\label{Ris:SZhV:uv}
\end{figure}

The following figures show the results of computations in a square and in a rectangular domain.
Fig.\,\ref{Ris:SZhV:Phi_norma} shows the isolines for the potential $\overline{\varphi}(x,y)$ with the step $0.05$.
Fig.\,\ref{Ris:SZhV:psi} demonstrates the streamlines of $\overline{\psi}(x,y,t)$ with the step $0.002$ at the instants
$t=10$ ($\thickapprox 0.78~s$) è $t=30$ ($\thickapprox 2.34~s$).
The isolines of the velocity field $\overline{\vu}(x,y,t)$ at $t=30$ are given in Fig.\,\ref{Ris:SZhV:uv}. After $t=30$
the flow is practically steady; for the additional control of the relaxation to a steady state we calculate the
mean-square norm  $\|\overline{\psi}(\cdot,t)\|$ (Fig.\,\ref{Ris:SZhV:Phi_norma}).
More detailed discussion of the computational results is given in Sect.\,\ref{i:7.0}. Here we just mention that Fig.\,\ref{Ris:SZhV:psi} shows the initial appearance of two co-rotating vortices. Later on, these two vortices merge into
a single vortex that represents an almost steady rotating flow in the whole domain. For the considered parameters the
transition (relaxation) to the final steady flow takes around $2~s$.

\begin{figure}[H]
\centering
\includegraphics[scale=0.40]{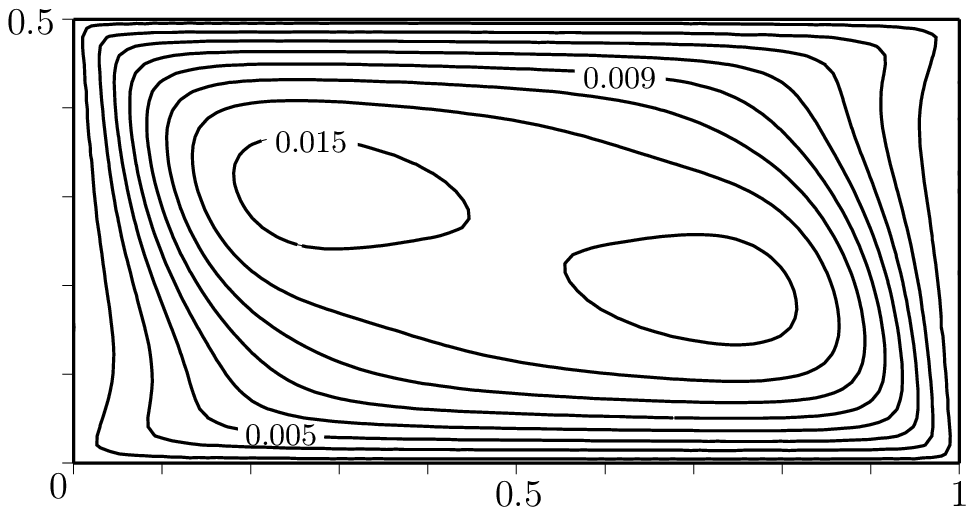}\
\includegraphics[scale=0.40]{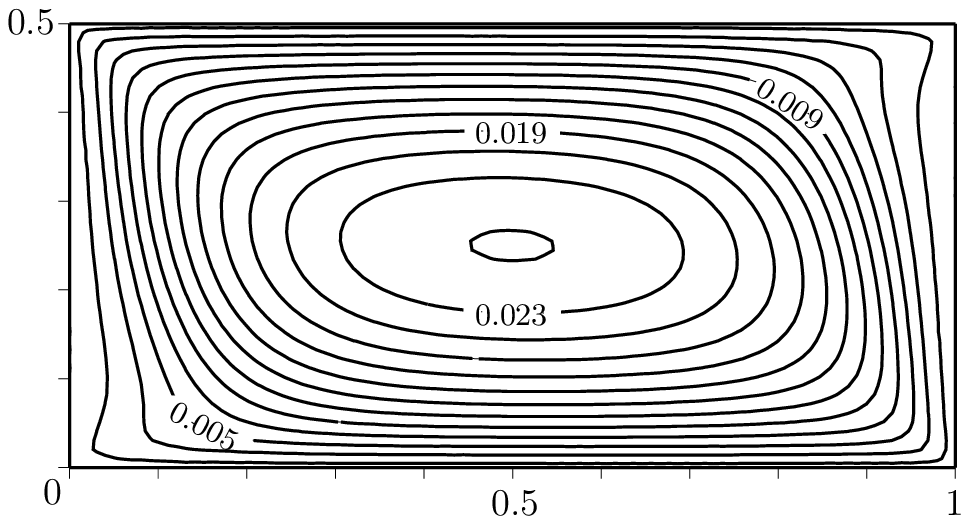}\\
\caption{The streamlines of $\overline{\psi}(x,y,t)$ for $X=1$, $Y=0.5$ at $t=7$ ($\thickapprox 0.546~s$) and
$t=20$ ($\thickapprox 1.56~s$)}\label{Ris:SZhV:psi21}
\end{figure}

\begin{figure}[H]
\centering  \includegraphics[scale=0.49]{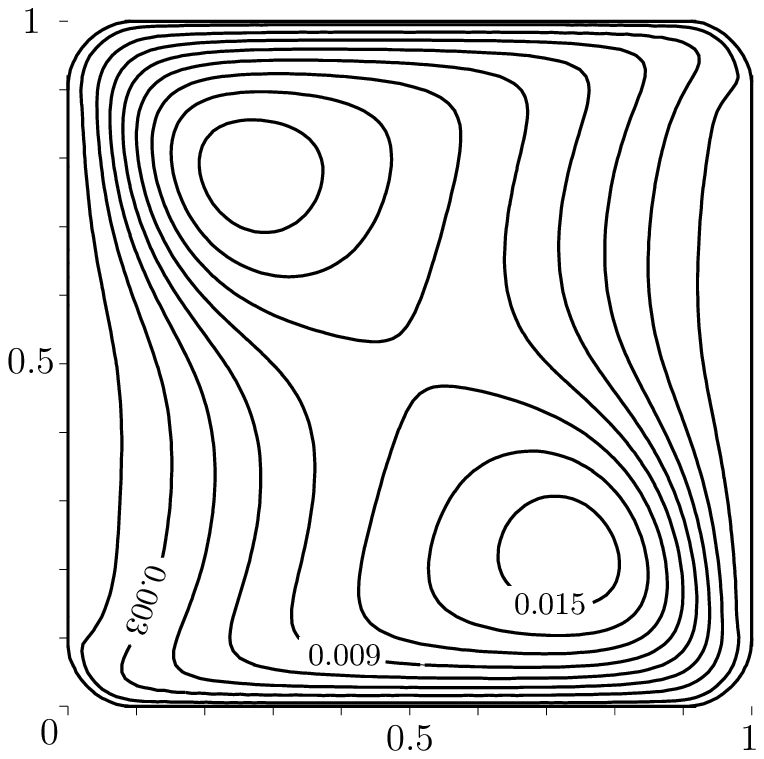}\quad
\centering  \includegraphics[scale=0.49]{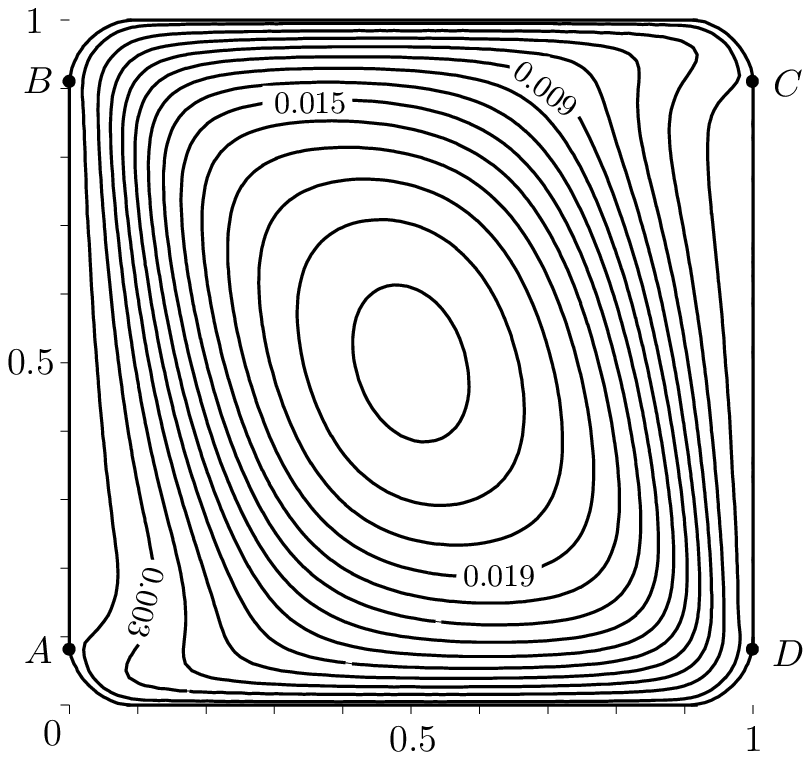}\quad
\caption{The streamlines for $\overline{\psi}(x,y,t)$ at $t=10$ ($\thickapprox 0.78~s$) and $t=30$ ($\thickapprox 2.34~s$)}
\label{Ris:SZhV:psi_round}
\end{figure}

In addition to the computations in a square domain, we perform the computations in rectangular domains with different
$Y$. In all cases $X > Y$ we observe a flow structure similar to the shown in Fig.\,\ref{Ris:SZhV:psi}: the initial
appearance of two vortices with the subsequent forming of an unified steady rotating flow. For example, the flow for
$X=1$, $Y=0.5$ at the instants $t=7$ ($\thickapprox 0.546~s$) è $t=30$ ($\thickapprox 2.34~s$) is shown in
Fig.\,\ref{Ris:SZhV:psi21}.
Fig.\,\ref{Ris:SZhV:psi_round} shows the flows for the square domain with the deliberately smoothed angles (the
curvature radius is $0.1$). One can see that the singularities in the electrical field near the vertices do not alter
the flow structure. In these computations we keep the boundary conditions (\ref{SZhV:eq:42a})--(\ref{SZhV:eq:44a}) at
$x=0,X$ the same, while on the rest of the boundary we introduce physically similar conditions. In these computations
$\overline{\varphi}=0$ on the part  $[A,B]$ of the boundary, and $\overline{\varphi}=\varphi_0$ on $[C,D]$. The
external electric field acts in the $y$-direction. On the rest of the boundary the tangential velocity component is
proportional to the tangential derivative of the potential (similar to (\ref{SZhV:eq:43a})). We also keep the no-leak
condition valid on the whole boundary.

\begin{figure}[H]
\centering  \includegraphics[scale=0.50]{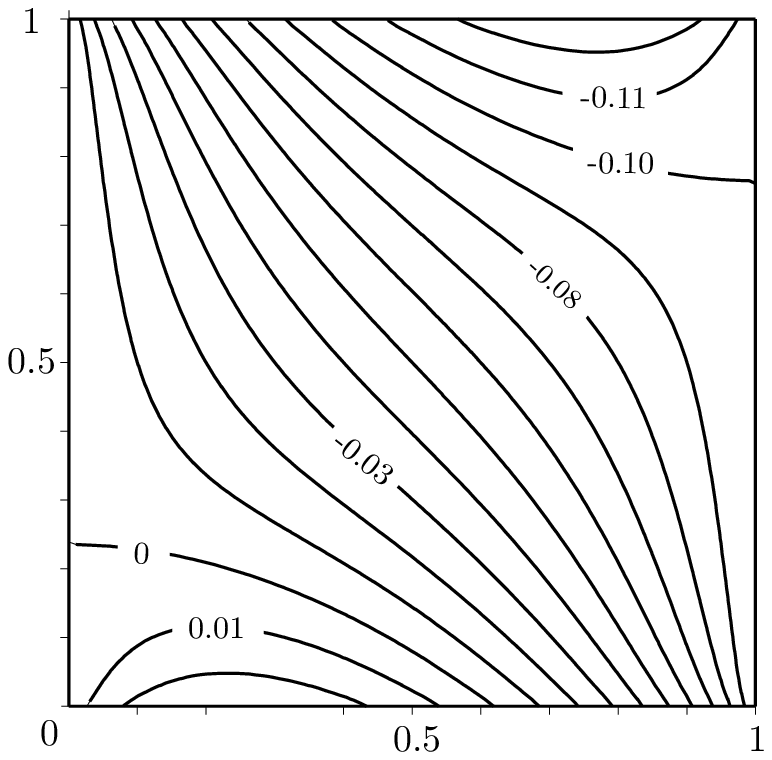}\quad
\centering  \includegraphics[scale=0.50]{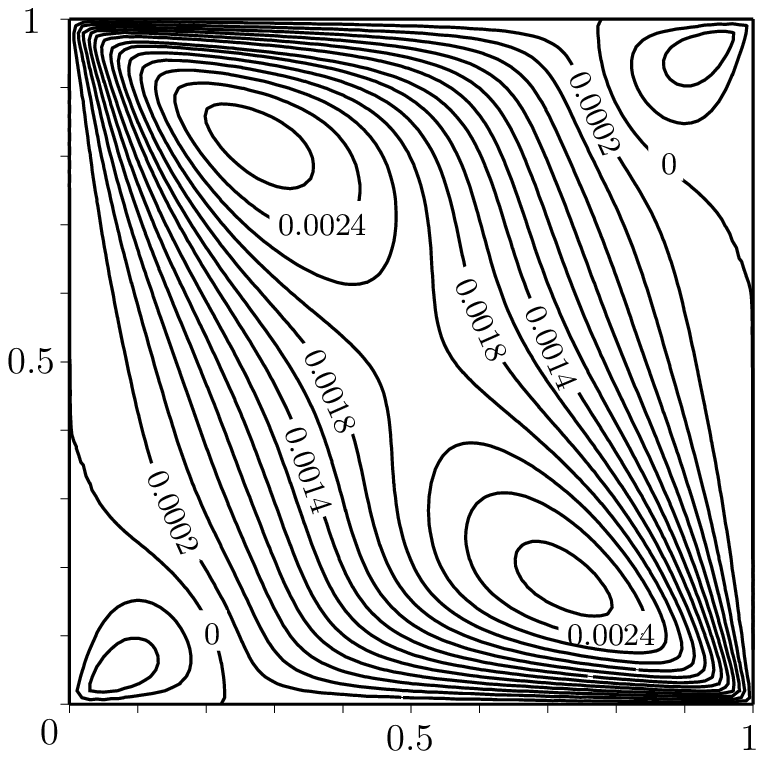}\\
\caption{The isolines of the potential $\overline{\varphi}(x,y)$ (left)
and the streamlines for $\overline{\psi}(x,y,t)$ at  $t=200$ ($\thickapprox 15.6~s$)}\label{Ris:SZhV:9}
\end{figure}

We have already mentioned that the tangential velocity at the boundary is determined by the relation between the
parameters $\varphi_0$, $E_0$, $X$, $Y$ (see (\ref{SZhV:eq:45a})) with one possible flow regime shown in
Fig.\,\ref{Ris:SZhV:3}. In order to confirm its existence we present in Fig.\,\ref{Ris:SZhV:9}  the results for the
parameters: $\varphi_0=-0.1$; $E_0=0.19$; $X=1$; $Y=1$. One can see there the isolines of the potential with the step
$0.01$ and the streamlines at $t=200$ with the step $0.0002$. The shown flow regime is almost steady: the norm
$\|\overline{\psi}(\cdot,t)\|=0.001442$ in the interval $160<t<200$ changes only in the last digit.
In Fig.\,\ref{Ris:SZhV:9} the tangential velocity at the boundary $y=0$ changes its sign at $x=X_0\approx 0.2$. The
computations show that the additional vortices in the angles of the domain do not appear if $X_0\lesssim 0.1$. In
particular, for $X=1$, $Y=1$, $E_0=0.19$ the generation of the rotating flow takes place when $|\varphi_0|>0.6$. It
also interesting to see the differences between the distributions of potentials (\emph{cf.}
Figs.\,\ref{Ris:SZhV:Phi_norma} and
\ref{Ris:SZhV:9}).

\section{Discussion}\label{i:7.0}

1. The existence of the discovered EHD rotational flow may be expected since it can be generated by the tangential
velocity at the boundaries. Nevertheless we should emphasise once more  that a rotational flow appears as the result of
the applying of \emph{\textbf{constant}} fields $\vE_{\text{out}}$ and $\varphi_0$, as it appears in the experiments
\cite{Amjadi01,Shirsavar02}.

2. The important result of this paper is the obtained in Sect.\,\ref{i:4.0} relation between the tangential velocity at
the boundary and Reynolds stresses. Our averaged equations (\ref{SZhV:eq:16a})--(\ref{SZhV:eq:20a}) are almost
identical to the derived in~\cite{Santiago_001,Santiago_002,Santiago_003,Santiago_004,Santiago_005}, although we used
different boundary conditions (\ref{SZhV:eq:8a})--(\ref{SZhV:eq:11a}). In
\cite{Santiago_001,Santiago_002,Santiago_003,Santiago_004,Santiago_005}  electrokinetic instability for the
solutions corresponding to inhomogeneous conductivity were studied. The Reynolds stresses terms were also  derived in
these papers, however they had been neglected due to their smallness. In our model
(\ref{SZhV:eq:16a})--(\ref{SZhV:eq:20a}) the situation is right the opposite. Reynolds stresses represent the main
reason for the appearance of the tangential velocity near the boundaries. One  can also see in
Appendix~\ref{app:sec:00} that our averaging method is more detailed than the one given
in~\cite{Santiago_001,Santiago_002,Santiago_003,Santiago_004,Santiago_005}.

3. A full quantitative comparison of our results with the experiments \cite{Amjadi01,Shirsavar02} is impossible, since
the key information about the values of some crucial parameters (\emph{e.g.} about the thickness of a film) is absent
in these papers.

4. The qualitative comparison of our results (Figs.\,\ref{Ris:SZhV:psi}, \ref{Ris:SZhV:psi21},
\ref{Ris:SZhV:psi_round}) with the flow pictures in \cite{Amjadi01,Shirsavar02} shows a good agreement:
both in the experiments and in our computations one can observe the appearance of the rotational flow, growing to its
stationary state during the time-interval of the order of 2~s. This fact opens the opportunity for a fast switching
between the directions of a rotation as has been proposed in~\cite{Amjadi01,Shirsavar02}. The magnitudes of rotational
velocities in our results and in the experiments are similar (around $3\,\textrm{cm}/\textrm{s}$, at least near the
boundaries).

5. In the experiments the flows with one vortex and with two vortices can be observed. Our computations show that only
one steady vortex can exist. Our results show (similar to the experiments) that there are two co-rotating vortices in
the rectangular film with the ratio of sizes 1:2. However our computations also show that such a flow is not steady, it
finally transforms to the flow with single vortex (Fig.\,\ref{Ris:SZhV:psi21}). However the authors
\cite{Amjadi01,Shirsavar02} do not mention whether or not the observed flow with two vortices is steady.
This contradiction can be resolved provided that the experimental observations correspond to an unsteady flow.

6. The experimental rotating flow
\cite{Amjadi01,Shirsavar02} appears only for some critical values of the electrical field $E_0^*$, which depend on
$\varphi_0^*$. The authors
\cite{Amjadi01,Shirsavar02} mistakenly stated that $E_0^*\varphi_0^*=\const$.
Their graph of this function in two logarithmic scales indeed represents a straight line, however its slope is not
$-1$. For our model (\ref{SZhV:eq:40a})--(\ref{SZhV:eq:44a}) a rotational flow also appears only for the certain values
of parameters. The rough estimation of these parameters follows from (\ref{SZhV:eq:45a}) (see also the comments to
Fig.\,\ref{Ris:SZhV:9}). The rotational flow with one vortex appears when the tangential velocity changes its sign at
the point $x\lesssim 0.1X$.

7. The experimental speed of the rotation does not depend on the viscosity $\nu^*$, while the formula
(\ref{SZhV:eq:46a}) for the tangential velocity gives $(a/\mathcal{T})\Rcf_3 \sim (\nu^*)^{-3}$. However for the
liquids with different viscosities (the solutions of glycerin in water) the thickness of the films also can be
different, while the velocity is $(a/\mathcal{T})\Rcf_3\sim (\nu^*)^{-3}h^4$. We are unable to compare this formula
with the experiments, since the data on a film thickness in \cite{Amjadi01,Shirsavar02} are absent.

8. In our model (\ref{SZhV:eq:16a})--(\ref{SZhV:eq:20a}) and in the numerical results
(Figs.\,\ref{Ris:SZhV:Phi_norma}--\ref{Ris:SZhV:psi_round}) the speed of the rotation decreases towards the center of a
film. It looks natural, since the cause of this rotation is the tangential velocity at the boundary  (see the boundary
conditions (\ref{SZhV:eq:42a})). In contrary, the results \cite{Amjadi01,Shirsavar02} show that the speed of the
rotation increases towards the center of a film. On the basis of this fact the authors of
\cite{Amjadi01,Shirsavar02} deny electrokinetic effects at the film edges as the possible mechanism that causes the
rotation. However one can propose a number of possible explanations for this discrepancy. First, it can be the
incompleteness of our mathematical model that does not consider the surface tension and the deviations of the free
surfaces of a film from the planes. Second, our mathematical model describes the \emph{averaged} velocity field that
differs from the real three-dimensional velocity distribution (see (\ref{SZhV:eq:A17})). Due to the accepted
electroneutrality of the mixture (almost everywhere except in the vicinities of the boundaries) the taking into account
the three-dimensionality of a flow can produce the decreasing of the rotation for the layers of a film near its
boundary. At the same time it is unclear whether the data in
\cite{Amjadi01,Shirsavar02} represent the average rotation speed or the speed of the rotation of the layer (\emph{e.g.} the
free surface) of a film. Third, a more complete mathematical model has to consider the Joule heat that naturally
appears in a weakly conductive liquid under a significant electrical current ($0.2\,\text{mA}\times 20\,
\text{V}=4\,\text{mW}$). The resulting nonuniform temperature can cause strong inhomogeneity in viscosity
and the permittivity of a solution. We should recall here that the changing of temperature in the interval
15--35$\,^\circ$C produces the changing of water permittivity $\varepsilon_r$ in the interval 81.9--74.8 ($\partial
\varepsilon_r/\partial T\approx 0.35$). For a strong electrical field it can produce a significant pondermotive force
$(1/2)\nabla\varepsilon (\nabla \varphi)^2$.

9. Our model of a rotational flow looks more realistic then the heuristic hypothesis of
\cite{Amjadi01,Shirsavar02} on the changing of the orientations of water molecular dipoles by an external electrical
field.

10. The rotating flow in our model is caused by
the tangential velocity applied at
the boundaries. This velocity has opposite directions at the different parts of the boundary. Therefore it is
interesting to study more systematically the vortex flows that appear at various critical values of the applied
tangential velocity.

11. Our model (\ref{SZhV:eq:16a})--(\ref{SZhV:eq:27a}) represents only a simplest asymptotic model of the flow near the
boundary. There is a serious potential for the development of this theory.  Here one should keep in mind that the
modelling of EHD processes in micro-scales represents a rather complex problem due to the broad spectrum of various
physical phenomena such as electrokinetic effects (electroosmosis, electrophoresis,
\emph{etc.}), the effects of diffusion, the chemical reactions both in a solution and on electrodes, the mass-transfer
by an electric field, the Joule heat, convection, Taylor-Aris dispersion,
\emph{etc.} In particular, it is unclear weather we can
consider the equilibrium Boltzmann concentrations $c_B\approx 10^{-4}\,\text{mol}/\text{m}^3$ or we have ions of only
one sign near the boundaries.

12. It is especially important to explain the connection between our model and the EDL-theories for strong external
electrical fields
\cite{Bazant_004,Bazant_005,Bazant_006,Ajdari_003,Ajdari_004,Zaltzman_Rubinstein,Dukhin}. In our model the rotating
flow is caused by the edge effects at the boundaries $y=0,\ Y$, where simplified boundary conditions lead to the
estimation of the value of $\Rcf = \Rcf_1(E_0) + \Rcf_3(E_0)$ (\ref{SZhV:eq:43a}), (\ref{SZhV:eq:46a}). At the same
time this simplified model can be upgraded with the use of contemporary EDL-theories (see also the references on
pp.\,\pageref{pageEDL1},
\pageref{pageEDL2}). This rather complex task can be undertaken if the industrial applications of
the \emph{liquid film motor flows} appear. Here one can go ahead with the full solution of the problem that must
include the exact evaluation of $\Rcf_1(E_0)$ and $\Rcf_3(E_0)$ and the correction of the assumption $E=\const$ in
(\ref{SZhV:eq:29a}) (see our remark on p.\,\pageref{page11}). To achieve such a goal one should describe an interface
flow more precisely, which is possible only with the use of EDL-theories. In general, the creation of a full industrial
level model requires to reconsider or upgrade all results of Sect.\,\ref{i:4.0}.

13. In practical applications the \emph{liquid film motor flows} can be used for the micromixing in microfluidic
devices.

14. The general significance of our results for the further developments of microhydrodynamics may consist in the
revaluation of the role of the considered classical effects in the micro- and nano-scale processes.

\begin{acknowledgments}
This research is partially supported by EPSRC (research grants GR/S96616/01, EP/D055261/1, and EP/D035635/1),
by the Russian Ministry of Education (programme `Development of the research potential
of the high school',  grants 2.1.1/6095 and 2.1.1/554), and by Russian Foundation for Basic Research (grants
07-01-00389, 08-01-00895, and 07-01-92213 NCNIL). The authors are grateful to the Department of Mathematics of the
University of York for the providing of excellent conditions for this research.
\end{acknowledgments}

\appendix

\section{The averaging procedure}\label{app:sec:00}

The averaging of (\ref{SZhV:eq:1a}), (\ref{SZhV:eq:3a})--(\ref{SZhV:eq:6a}), that takes into account the boundary
conditions (\ref{SZhV:eq:8a})--(\ref{SZhV:eq:11a}), gives the exact but not closed system of equations
\begin{eqnarray}
\displaystyle
 \delta^2 (\partial_t\overline{\vu}+
 \overline{\vu} \cdot \nabla_0 \overline{\vu} )+\delta^2 \Div_0 ( \overline{\widetilde{\vu}\otimes\widetilde{\vu}})
      =  \nonumber \\
=  \displaystyle -        \delta^2\nabla_0 \overline{p}+
      \delta^2\nu
        \Delta_0 \overline{\vu}
      - \overline{q}\nabla_0 \overline{\varphi}-
      \overline{\widetilde{q}\nabla_0 \widetilde{\varphi}},
\label{SZhV:eq:A1}
\end{eqnarray}
\begin{equation}\label{SZhV:eq:A2}
  \Div_0 \overline{\vu} =0,
\end{equation}
\begin{equation}\label{SZhV:eq:A3}
   \varepsilon \Delta_0 \overline{\varphi}=- \overline{q},
\end{equation}
\begin{equation}\label{SZhV:eq:A4}
 \partial_t\overline{c}_k+
 \overline{\vu} \cdot \nabla_0 \overline{c}_k +
 \Div_0 (\overline{\widetilde{\vu}\widetilde{c}_k})
+ D_k \Div_0 \overline{\vi}_k=0,
\end{equation}
\begin{equation}\label{SZhV:eq:A5}
\overline{\vi}_k = -D_k(\nabla_0 \overline{c}_k + \zk \gamma \overline{c}_k \nabla_0 \overline{\varphi}+
 \zk \gamma \overline{\widetilde{c}_k \nabla_0 \widetilde{\varphi}}).
\end{equation}
In order to obtain the closed system with the precision $O(\delta^4)$  we use the decompositions (\ref{SZhV:eq:15a}) to
calculate the terms
\begin{equation}\label{SZhV:eq:A6}
\overline{\widetilde{\vu}\otimes\widetilde{\vu}}=
\overline{\widetilde{\vu}^0\otimes\widetilde{\vu}^0}+O(\delta^2),
\end{equation}
\begin{equation}\label{SZhV:eq:A7}
 \overline{\widetilde{q}\nabla_0 \widetilde{\varphi}}=\overline{(\widetilde{q}^0+\delta^2\widetilde{q}^1)\nabla_0 (\widetilde{\varphi}^0+\delta^2\widetilde{\varphi}^1)}+
O(\delta^4),
\end{equation}
\begin{equation}\label{SZhV:eq:A8}
\overline{\widetilde{\vu} \widetilde{c}_k}=
\overline{(\widetilde{\vu}^0+\delta^2\widetilde{\vu}^1) (\widetilde{c}_k^0+\delta^2\widetilde{c}_k^1)}
+O(\delta^4),
\end{equation}
\begin{equation}\label{SZhV:eq:A9}
\overline{\widetilde{c}_k \nabla_0 \widetilde{\varphi}}=
\overline{(\widetilde{c}_k^0+\delta^2\widetilde{c}_k^1) \nabla_0 (\widetilde{\varphi}^0+\delta^2\widetilde{\varphi}^1)}
+O(\delta^4).
\end{equation}
For the main terms in (\ref{SZhV:eq:15a}) equations (\ref{SZhV:eq:2a}), (\ref{SZhV:eq:4a})--(\ref{SZhV:eq:6a}) and
condition (\ref{SZhV:eq:10a}) yield
\begin{equation}\label{SZhV:eq:A10}
(\overline{q}^0+\widetilde{q}^0)\partial_z \widetilde{\varphi}^0=0, \quad   \overline{q}^0+\widetilde{q}^0 = \sum_{k}\zk (\overline{c}_k^0+\widetilde{c}_k^0),
\end{equation}
\begin{equation}\label{SZhV:eq:A11}
\partial_z \widetilde{I}^0_k =0, \quad  \overline{I}^0_k+\widetilde{I}^0_k =\partial_z \widetilde{c}^0_k + \zk \gamma  (\overline{c}^0_k+\widetilde{c}^0_k) \partial_z \widetilde{\varphi}^0.
\end{equation}
\begin{equation*}
  (\overline{I}^0_k+\widetilde{I}^0_k)\bigl|_{z=\pm1} = 0 .
\end{equation*}
Equations (\ref{SZhV:eq:A10}), (\ref{SZhV:eq:A11}) give
 $\partial_z\widetilde{\varphi}^0=0$,
 $\partial_z\widetilde{I}_k^0=0$,
 $\widetilde{I}_k^0=\partial_z\widetilde{c}_k^0$. It is clear that if
 $\partial_z\widetilde{f}=0$, then $\widetilde{f}=0$ and
 $f=\overline{f}$. Hence
\begin{equation}\label{SZhV:eq:A12}
\widetilde{\varphi}^0 =0, \quad \widetilde{c}^0_k=0, \quad \widetilde{q}^0=0,\\
\end{equation}
\begin{equation*}
\varphi^0=\overline{\varphi}^0,
\quad c_k^0=\overline{c}_k^0,
\quad q^0=\overline{q}^0 .
\end{equation*}
The use of (\ref{SZhV:eq:A12}) transforms the expressions (\ref{SZhV:eq:A7})--(\ref{SZhV:eq:A9})  to the form
\begin{equation}\label{SZhV:eq:A13}
 \overline{\widetilde{q}\nabla_0 \widetilde{\varphi}}=
O(\delta^4),
\end{equation}
\begin{equation*}
\overline{\widetilde{\vu} \widetilde{c}_k}=
\delta^2\overline{\widetilde{\vu}^0\widetilde{c}_k^1}
+O(\delta^4),\quad \overline{\widetilde{c}_k \nabla_0 \widetilde{\varphi}}=
O(\delta^4).
\end{equation*}
From (\ref{SZhV:eq:1a}), (\ref{SZhV:eq:A12}) we obtain the equation for $\widetilde{\vu}^0$
\begin{equation*}
  \nu
       \partial_{zz}\widetilde{\vu}^0
      - \overline{q}^0\nabla_0 \overline{\varphi}^0=0,
\end{equation*}
which is required for the calculation of (\ref{SZhV:eq:A6}) with the precision $O(\delta^2)$. In particular it means
that we can make the replacements $\overline{q}=\overline{q}^0+O(\delta^2)$,
$\overline{\varphi}=\overline{\varphi}^0+O(\delta^2)$ and $\widetilde{w}_0$, $\widetilde{\vu}_0$ can be found from the
equations
\begin{equation}\label{SZhV:eq:A14}
  \nu
       \partial_{zz}\widetilde{\vu}^0
      - \overline{q}\nabla_0 \overline{\varphi}=0,
\end{equation}
\begin{equation}\label{SZhV:eq:A15}
\Div_0 (\overline{\vu}^0+\widetilde{\vu}^0) +  \partial_z\widetilde{ w}^0 =0,
\end{equation}
with the boundary condition
\begin{equation}\label{SZhV:eq:A16}
\overline{w}^0+\widetilde{w}^0=0, \quad z=\pm1.
\end{equation}
We assume that $\overline{w}^0=0$. The integration of (\ref{SZhV:eq:A14})--(\ref{SZhV:eq:A16}) yields
\begin{equation}\label{SZhV:eq:A17}
 \widetilde{\vu}^0 = g'(z)\vU,\ \
 \widetilde{w}^0 = -g(z)\Div_0 \vU,\ \
 \nu\vU=\overline{q}\nabla_0 \overline{\varphi},
\end{equation}
\begin{equation*}
 \Div_0 \overline{\vu}^0=0,\ \ g(z)=\frac16(z^3-z),\ \
 \overline{g(z)}=0,\ \  \overline{g'(z)}=0,
\end{equation*}
where we have used the notation (\ref{SZhV:eq:20a}) for $\vU$.

One can notice that we do not require $\widetilde{\vu}^0$ to satisfy the boundary condition (\ref{SZhV:eq:9a}). This
condition is required only for $\widetilde{\vu}$. The equality $\partial_z\widetilde{\vu}^0=0$ at $z=\pm1$ leads  to
$\vU=0$ that is not true. In the exact problem one should consider a boundary-layer solution at $z=\pm1$ and assume the
absence of the charge ($\overline{q}=0$) at the boundary. In the opposite case the action of a tangential to the
boundary external field creates the stresses related to Maxwell's electromagnetic stress tensor.

The use of (\ref{SZhV:eq:A17}) gives the expression for (\ref{SZhV:eq:A6})
\begin{equation}\label{SZhV:eq:A18}
  \overline{\widetilde{\vu}\otimes\widetilde{\vu}}=
\overline{g'^2(z)}( \vU\otimes\vU)+O(\delta^2),\quad
\overline{g'^2(z)}=\frac1{45}.
\end{equation}
The calculation of $\overline{\widetilde{\vu} \widetilde{c}_k}$ is based on the next approximation for the equations
(\ref{SZhV:eq:4a})--(\ref{SZhV:eq:6a})
\begin{equation*}
 \partial_t c_k^0+\vu^0\cdot \nabla_0 c_k^0 +
 w^0\partial_z c_k^0 + \Div_0 \vi_k^0 +
 \partial_z I_k^1=0,
 \end{equation*}
\begin{equation*}
 \vi_k^0 = -D_k\Bigl(\nabla_0 c_k^0+\zk\gamma c_k^0\nabla_0\varphi^0 \Bigr),
 \end{equation*}
\begin{equation*}
  I_k^1 = -D_k\Bigl(\partial_z c_k^1+\zk\gamma( c_k^1\partial_z\varphi^0+
 c_k^0\partial_z\varphi^1) \Bigr),
\end{equation*}
\begin{equation*}
    \varepsilon(\Delta_0 \varphi^0 +
  \partial_{zz}\varphi^1) =- q^0
\end{equation*}
or taking in account (\ref{SZhV:eq:A12})
\begin{equation}\label{SZhV:eq:A19}
 \partial_t \overline{c}_k^0+(\overline{\vu}^0+\widetilde{\vu}^0)\cdot \nabla_0 \overline{c}_k^0 +
 \Div_0 \vi_k^0 +
 \partial_z I_k^1=0,
 \end{equation}
\begin{equation*}
 \vi_k^0 = -D_k\Bigl(\nabla_0 \overline{c}_k^0+\zk\gamma \overline{c}_k^0\nabla_0\overline{\varphi}^0 \Bigr),
 \end{equation*}
\begin{equation*}
  I_k^1 = -D_k\Bigl(\partial_z \widetilde{c}_k^1+\zk\gamma  \overline{c}_k^0\partial_z\widetilde{\varphi}^1 \Bigr),
\end{equation*}
\begin{equation*}
    \varepsilon(\Delta_0 \overline{\varphi}^0 +
  \partial_{zz}\widetilde{\varphi}^1) =- \overline{q}^0.
\end{equation*}
The last equation shows that $\partial_{zz}\widetilde{\varphi}^1$ does not depend on $z$, so to find
$\widetilde{c}_k^1$ we obtain the equation
\begin{equation}\label{SZhV:eq:A20}
 \widetilde{\vu}^0\cdot \nabla_0 \overline{c}_k^0 -D_k
 \partial_{zz}\widetilde{c}_k^1=0
\end{equation}
with the boundary conditions that follow from  (\ref{SZhV:eq:10a})
\begin{equation*}
 \partial_z \widetilde{c}_k^1\bigl|_{z=\pm1}=0.
\end{equation*}
For the calculation of  $\overline{\widetilde{\vu} \widetilde{c}_k}$ one can take
$\overline{c}_k=\overline{c}_k^0+O(\delta^2)$ in (\ref{SZhV:eq:A20}), since the required precision for
(\ref{SZhV:eq:A8}) is $O(\delta^2)$. It allows us to integrate the equation (\ref{SZhV:eq:A20})
\begin{equation*}
 D_k
 \widetilde{c}_k^1=\bigl(g_0(z)-\overline{g_0(z)}\bigr)\vU \cdot \nabla_0 \overline{c}_k,\
 g_0(z) = \frac{z^2}{12}\left(\frac12z^2 -1 \right).
\end{equation*}
Finally we obtain
\begin{equation}\label{SZhV:eq:A21}
\overline{\widetilde{\vu}^0 \widetilde{c}_k^1} =-
\alpha_k\vU(\vU \cdot \nabla_0 \overline{c}_k),
\end{equation}
\begin{equation*}
\alpha_k=-
\frac{1}{D_k}\left(\overline{g'(z)\bigl(g_0(z)-\overline{g_0(z)}\bigr)}\right)
=\frac{4}{945D_k}.
\end{equation*}

\section{}\label{app:sec:01}

Let us show that in the case (\ref{SZhV:eq:29a}) equations (\ref{SZhV:eq:16a}) can be integrated. The use
(\ref{SZhV:eq:36a}) and (\ref{SZhV:eq:20a}) gives the velocity component $\overline{u}$
\begin{equation}\label{SZhV:eq:B1}
 \frac{\beta}{\nu^2}
E \partial_y \left(
 \overline{q}^2  \partial_y \Phi\right)=-\delta^2 \partial_x
 p+\nu \delta^2 \partial_{yy} \overline{u}-
 \overline{q} E.
\end{equation}
Taking in account that $\overline{q}$, $\Phi$ and $\overline{u}$ depend only on $y$ we get
\begin{equation*}
 -\delta^2 p =
 \left\{
 \frac{\beta}{\nu^2} E \partial_y \left(
 \overline{q}^2  \partial_y \Phi\right)
 - \nu \delta^2 \partial_{yy} \overline{u} +
 \overline{q} E
 \right\}x + H(y).
\end{equation*}
Its substitution into the equation (\ref{SZhV:eq:16a}) for $w$ shows that $p$ depends on $y$ only
\begin{equation*}
 \frac{\beta}{\nu^2} \partial_y \left(
 \overline{q}  \partial_y \Phi\right)^2 + \overline{q}  \partial_y \Phi= H'(y).
\end{equation*}
It follows that the expression in braces is vanishing and (\ref{SZhV:eq:B1}) gives (\ref{SZhV:eq:37a}).

\section{}\label{app:sec:02}

Let us consider the case when the values of all diffusion coefficients are the same ($D_k\equiv D$). The multiplying of
(\ref{SZhV:eq:A4}) by $\zk$ and combining the results  yield
\begin{equation*}
\frac{d\overline{q}}{dt}
+
 \Div_0 (\overline{\widetilde{\vu}\widetilde{q}})
 - D\Div_0 (\nabla_0 \overline{q}+\sigma \nabla_0 \overline{\varphi})=0,
\end{equation*}
where $\sigma=D \sum_{k} \zk^2 \gamma \overline{c}_k$ is the conductivity of a mixture;  we have also taken
(\ref{SZhV:eq:A21}) into account. By virtue of (\ref{SZhV:eq:A21}) the electroneutrality $\overline{q}=0$ leads to
$\widetilde{q}=0$ everywhere except the boundaries. Hence, in the case $\overline{c}_k=c_B$ (see (\ref{SZhV:eq:31a}))
we arrive to the equation (\ref{SZhV:eq:41a}).

Recall that the requirement of the equality of all diffusion coefficients represents a strong restriction. In
particular, the difference between the diffusion coefficients leads to the participation of the term
$\overline{\widetilde{\vu}\widetilde{q}}$ (linked to the Taylor--Aris dispersion
\cite{Santiago_001,Santiago_002,Santiago_003,Santiago_004,Santiago_005}) into the electrokinetic instabilities.

\end{document}